\documentclass[runningheads]{llncs}
\usepackage[T1]{fontenc}
\usepackage{graphicx}
\usepackage{hyperref}
\usepackage{adjustbox}
\usepackage{color}

\hypersetup{
    colorlinks=true,
    linkcolor=blue,
    filecolor=blue,      
    urlcolor=blue,
    pdftitle={Overleaf Example},
    pdfpagemode=FullScreen,
    citecolor=blue
    }
\usepackage{amsmath}
\usepackage{multirow}
\usepackage{graphicx}
\usepackage{subcaption}
\usepackage{subfiles} 
\usepackage{cleveref}
\usepackage{booktabs}
\usepackage[dvipsnames,table,xcdraw]{colortbl}
\usepackage{subfiles} 
\usepackage{tabu}
\usepackage{url}
\usepackage{bbding}
\usepackage{tablefootnote}
\begin{document}
\title{\texttt{NeuralOCT}: Airway OCT Analysis \\ via Neural Fields}

\author{Yining Jiao\textsuperscript{1}, Amy Oldenburg\textsuperscript{1}, Yinghan Xu\textsuperscript{1}, Srikamal Soundararajan\textsuperscript{1},  Carlton Zdanski\textsuperscript{1}, Julia Kimbell\textsuperscript{1}, Marc Niethammer\textsuperscript{1}\thanks{Corresponding author.}} 
\institute{\textsuperscript{1}University of North Carolina at Chapel Hill}

%
%
%
\maketitle              
\begin{abstract}
Optical coherence tomography (OCT)  is a popular modality in ophthalmology and is also used intravascularly. Our interest in this work is OCT in the context of airway abnormalities in infants and children where the high resolution of OCT and the fact that it is radiation-free is important. The goal of airway OCT is to provide accurate estimates of airway geometry (in 2D and 3D) to assess airway abnormalities such as subglottic stenosis. We propose \texttt{NeuralOCT}, a learning-based approach to process airway OCT images. Specifically, \texttt{NeuralOCT} extracts 3D geometries from OCT scans by robustly bridging two steps: point cloud extraction via 2D segmentation and 3D reconstruction from point clouds via neural fields. Our experiments show that \texttt{NeuralOCT} produces accurate and robust 3D airway reconstructions with an average A-line error smaller than 70 micrometer. Our will be  available on GitHub.
\end{abstract}
\section{Introduction}

\noindent{\bf OCT applications.} Optical coherence tomography (OCT) is based on laser light tissue interactions which are processed to allow tissue imaging. As OCT allows for high resolution imaging it is, for example, able to visualize small retinal lesions or plaque in blood vessels~\cite{drexler2008state,tearney2012consensus}.  
Recently airway OCT~\cite{price2018geometric,wijesundara2014quantitative} has gained popularity to assess airway geometry. Our work focuses on OCT to ultimately assess airway abnormalities in infants and children where OCT's high resolution and its ability to image without radiation are important considerations. 
Assessing airway abnormalities, such as subglottic stenosis which can severely impact airflow, is important to recommend either surgery or watchful-waiting (i.e., to see if an airway  will normalize through normal aging). 

\begin{figure}[tp]
\centering
\begin{subfigure}[t]{0.5\linewidth}
    \centering
    \includegraphics[width=\linewidth]{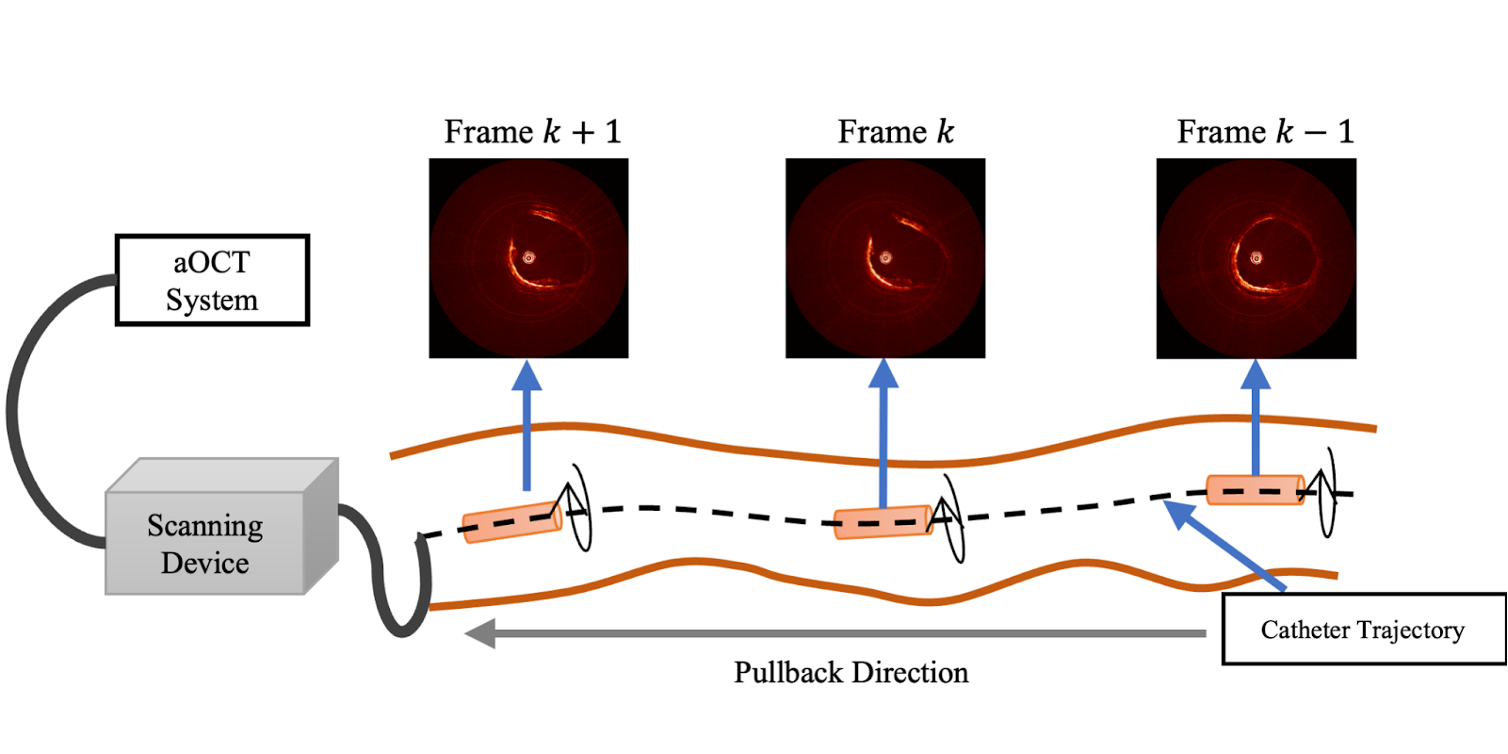}
    \caption{}
    \label{fig.aoct_scan_a}
\end{subfigure}\hfill
\begin{subfigure}[t]{0.24\linewidth}
    \centering
    \includegraphics[width=\linewidth]{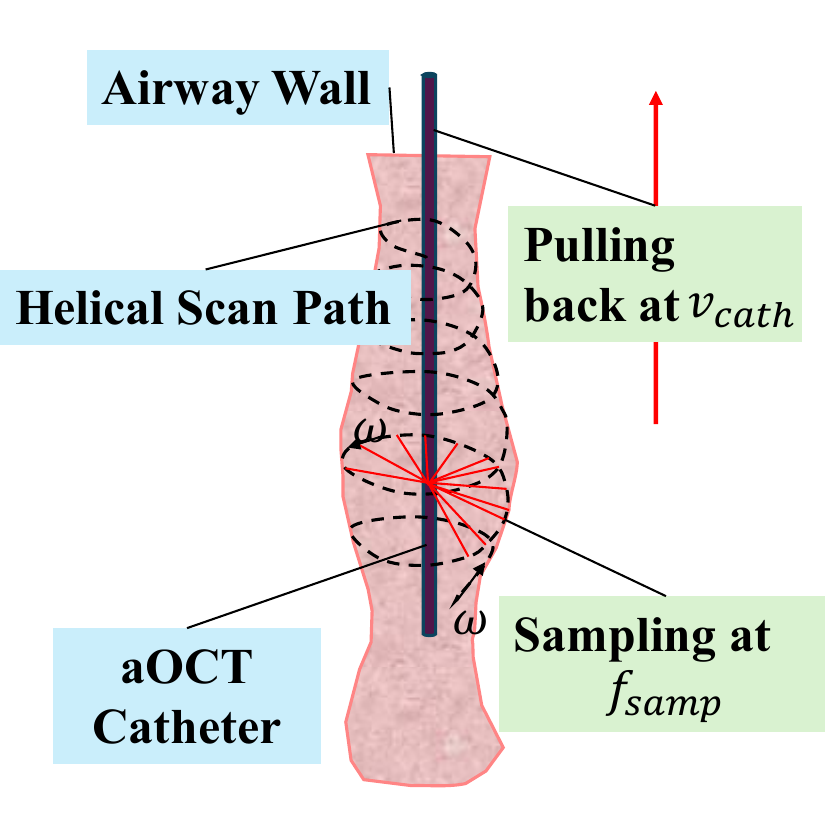}
    \caption{}
    \label{fig.aoct_scan_b}
 \end{subfigure}
 \begin{subfigure}[t]{0.24\linewidth}
    \centering
    \includegraphics[width=\linewidth]{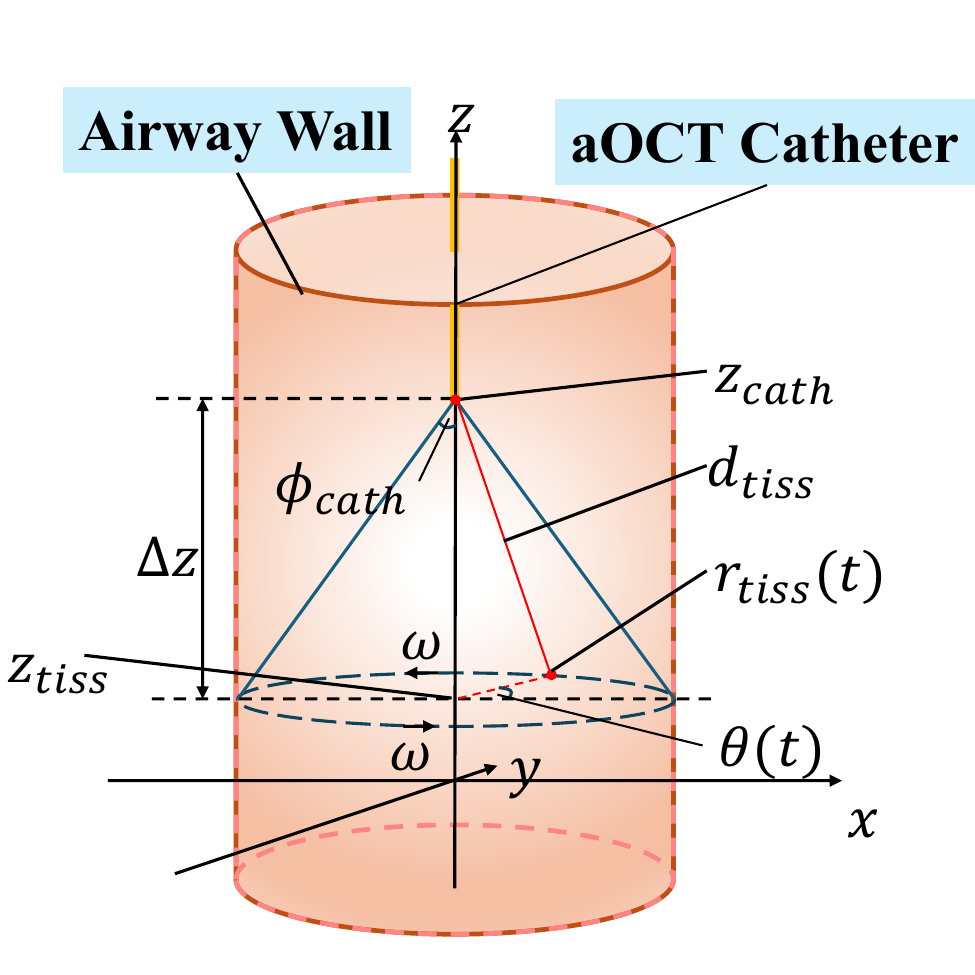}
    \caption{}
    \label{fig.aoct_scan_c}
 \end{subfigure}
     \caption{Principle of the aOCT scanning process. (a) describes the airway OCT scanning process, during which frames are captured as the catheter is pulled back from the bottom to the top of the airway. In (b), laser rays are emitted from the catheter and hit the airway wall to estimate line-of-sight distance as helical A-lines. (c) illustrates how to relate the coordinates on the airway wall to the catheter and laser geometry in a cylindrical coordinate system.}
    \label{fig.oct_scanning}
\end{figure}

\vskip1ex
\noindent{\bf OCT principle.} Unlike retinal OCT, airway OCT is endoscopic. Fig.~\ref{fig.aoct_scan_a} illustrates the principle behind airway OCT: the OCT catheter travels along the airway lumen and emits light through a rotating laser. The reflected light of one revolution is then used to reconstruct tissue information resulting in reconstructed 2D OCT frames. As the scanning speed is fast these 2D OCT frames approximate planar cuts through the airway. Different from volume-based imaging such as magnetic resonance imaging (MRI) or computed tomography (CT), one pull-back OCT scan for an airway is based on  a helical scan path with respect to the catheter trajectory. Every point on this path is associated with an A-line: an OCT depth-profile based on the emitted and reflected laser light along a ray. Therefore, one does not directly obtain a voxel grid or a mesh representation of anatomies of interest. Instead airway geometry needs to be reconstructed considering the helical path of an anatomic OCT (aOCT) pullback scan. Obtaining accurate airway geometry is important for downstream tasks, e.g., to facilitate airway shape analysis or simulated surgeries via shape editing. 

\vskip1ex
\noindent{\bf Motivation.} In this work, we propose \texttt{NeuralOCT}, a learning-based approach for geometry reconstruction from OCT. Specifically, \texttt{NeuralOCT} extracts 3D geometries from OCT scans by combining two steps: 1) point cloud extraction via 2D segmentation and 2) 3D reconstruction from point clouds via neural fields. Our \texttt{NeuralOCT} approach is based on a 2D segmentation network working as a teacher module to produce raw point clouds on the airway wall; the neural field is then the student receiving and filtering these 2D-segmentation-derived point clouds to produce a 3D geometry reconstruction at infinite resolution. 

\vskip1ex
The technical contributions of \texttt{NeuralOCT} and their clinical significance is: 
\begin{itemize}
    \item[1)] We investigate the suitability of advanced segmentation techniques for aOCT;
    \item[2)] \texttt{NeuralOCT} is the first approach to recover 3D geometries from raw point clouds obtained via 2D OCT segmentations;
    \item[3)] \texttt{NeuralOCT} is the first approach using neural fields to represent 3D geometries from OCT scans, which is expected to simplify further downstream tasks such as shape analysis and simulated surgeries.
\end{itemize}

\section{Related Work}

\noindent\textbf{Automatic Airway OCT Processing.} Several approaches for airway OCT processing have recently been developed. Zhuang et al.~\cite{zhuang2023octpipeline} developed an automatic 2D segmentation method based on dynamic programming and multiple filtering steps. These 2D segmentations are then concatenated along the catheter trajectory to obtain a 3D reconstruction. Only two OCT scans are used for method development and evaluation hence it remains unclear how well their method generalizes. Zhou et al.~\cite{zhou2023dlforoctseg} proposed using a 2D CNN segmentation model to measure endobronchial OCT parameters but did not target 3D geometries. \emph{\texttt{NeuralOCT} proposes a deep-learning based 3D-aware approach to estimate accurate 3D airway geometry from OCT scans.}

\vskip1ex
\noindent\textbf{Neural Fields.}
Compared to explicit geometry representations such as voxel grids~\cite{wu2016learning}, point clouds~\cite{achlioptas2018learning} and meshes~\cite{groueix2018papier}, neural fields represent geometry based on a function which is represented by a neural network. Neural fields are able to represent highly detailed and complex signals using a relatively small amount of data~\cite{park2019deepsdf,mescheder2019occupancy,jiao2023texttt}. The basic idea behind neural fields for shape representation is to replace grid-based parameterizations of level set functions~\cite{sethian1999level,osher2005level} (where a shape is a specific level set) by an actual functional representation which is parameterized via a deep neural network. Hence, neural fields are not reliant on meshes, grids, or a discrete set of points. This allows them to efficiently represent natural-looking surfaces~\cite{gropp2020implicit,sitzmann2020siren,niemeyer2019occupancy}. Further, neural fields can  estimate geometries from incomplete or noisy point clouds, for example, obtained from LiDAR observations~\cite{neuralpull,noise2noisemapping}. \emph{\texttt{NeuralOCT} is the first approach to use neural fields to extract and represent highly-detailed 3D airway geometries from OCT segmentations.}
 

\vskip1ex
\noindent\textbf{Medical Image Segmentation based on Limited Image Sample Sizes.}
Our airway OCT segmentation task is challenging due to sparse supervision and limited data availability. One solution to deal with these challenges is to finetune a pre-trained foundation model, such as MedSAM~\cite{ma2024medsam}, which itself is based on finetuning the Segment Anything Model (SAM)~\cite{kirillov2023sam} on large-scale medical image datasets. Another solution is to train a data-efficient segmentation model from scratch. The popular nnU-Net~\cite{isensee2021nnu} provides an out-of-the-box solution which automatically provides the best configuration for a given dataset. It  analyzes the provided training cases and automatically configures a suitable U-Net for segmentation. \emph{\texttt{NeuralOCT} investigates the suitability of foundation models and nnU-Net for airway OCT segmentation.}

\section{Background on aOCT}
\label{sec.prel}
This section introduces the mathematical formulation underlying anatomic OCT (aOCT). See ~\cite{bu2017airway,wijesundara2014quantitative} for further details on aOCT. Briefly, as illustrated in Fig.~\ref{fig.aoct_scan_a}, a fiber-optic catheter delivers the sample arm light (the red lines in Fig.~\ref{fig.aoct_scan_b}) and is set up to scan the airway wall helically by simultaneously rotating and translating (forming a pull-back scan)~\cite{price2018geometric}. As a result, the rays impinging on and penetrating the airway wall provide reconstructed A-lines which, for one laser revolution, result in a 2D aOCT image frame as shown in Fig.~\ref{fig.aoct_scan_a}.

Specifically, the catheter (and with it the laser) is pulled back from the bottom to the top of the airway at a speed of $v_{cath}$ while delivering the rotating laser light rays at an angular speed of $\omega$ which hit the airway wall and penetrate the tissue. The reflected light is used to reconstruct an A-line. At a certain time point $t$, the current rotating angle of the laser ray is $\theta(t)= \omega \cdot t$. 
The current position of the catheter (with the laser source) is $z_{cath}(t)= \int \mathbf{v}(t)~dt$, where $\mathbf{v}$ describes the velocity of the catheter as it is pulled back. The scanning frequency is $f_{samp}$, meaning that there will be $f_{samp}$ recorded A-lines (from which we will extract points on the airway wall) per second. 

Fig.~\ref{fig.aoct_scan_c} illustrates the geometric relationship of the key parameters and variables in the aOCT scanning process. The polar angle of the laser beam relative to the fiber optic axis is fixed to $\phi_{cath}$. The aOCT system measures the line-of-sight distance from the catheter to the airway wall, $d_{tiss}(t)$. Throughout time, the airway wall position illuminated by the laser can be described in cylindrical coordinates as: $(d_{tiss}(t), \theta(t), z_{cath}(t))$, in which $z_{cath}(t)$ represents the catheter position and $d_{tiss}(t)$ represents the line-of-sight-distance from the catheter to the airway wall. Since $\theta(t)$ is already known, we are able to use $d_{tiss}(t), z_{cath}(t)$ and $\phi_{cath}$ to derive $r_{tiss}$ and $z_{tiss}$ as follows:
\begin{align}
r_{tiss}(t) = d_{tiss}(t) \cdot sin({\phi}_{cath})\,,\\
z_{tiss}(t) = z_{cath}(t) - \Delta z(t) = z_{cath}(t) - d_{tiss}(t)cos(\phi_{cath})\,,
\label{eq.rtiss}
\end{align}
where $z_{tiss}(t)$ represents the position of the sampled point on the airway wall mapped on the fiber optic axis and $r_{tiss}(t)$ represents the sides of a right triangle with hypotenuse $d_{tiss}(t)$.

Based on this aOCT scanning process, we can directly obtain rectangular aOCT images as depicted in Fig.~\ref{fig.workflow}. 
The rectangular frames are in a polar coordinate system, where the $x$ axis represents the rotating angle $\theta(t)$, and the $y$ axis represents the light-of-sight distance from the airway wall to the catheter axis $d_{tiss}(t)$. The image intensities reflect how strongly the laser rays are reflected. From the rectangular aOCT frames in polar coordinates, we can reconstruct the corresponding polar aOCT frames (as shown in Fig.~\ref{fig.aoct_scan_a}) in Cartesian coordinates.

\section{Method}
\begin{figure}
\vspace{-0.1in}
    \centering
    \includegraphics[width=1.\linewidth]{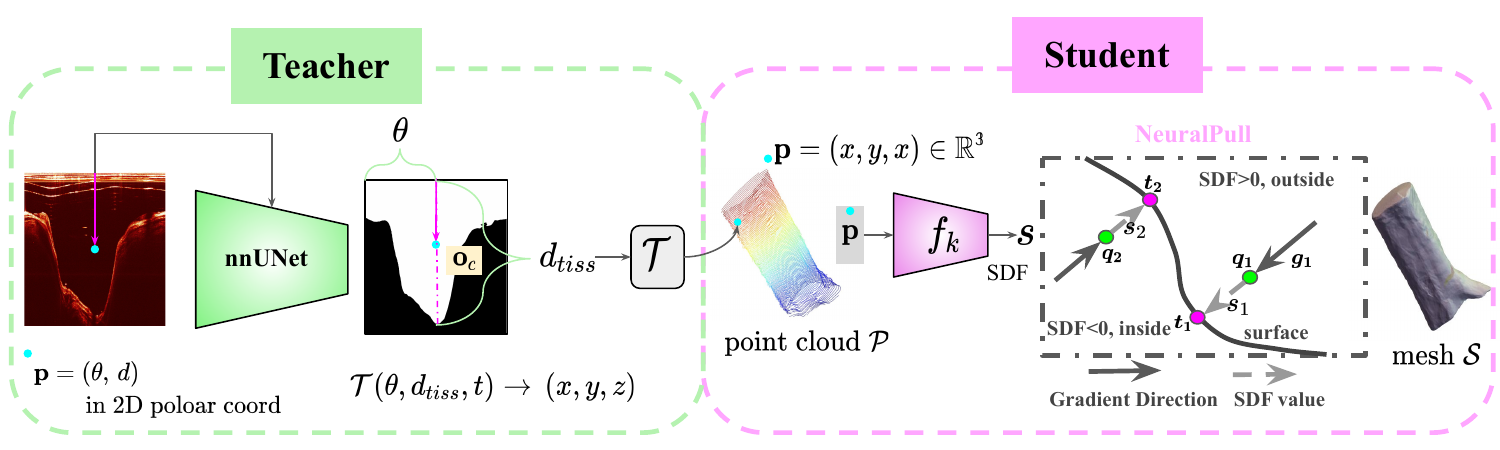}
    \caption{Principle of the \texttt{NeuralOCT} approach. \texttt{NeuralOCT} extracts 3D geometries from OCT scans by combining point cloud extraction from 2D segmentations with 3D reconstruction via neural fields.}
    \label{fig.workflow}
\end{figure}

Fig.~\ref{fig.workflow} provides an overview of how \texttt{NeuralOCT} extracts 3D geometries from OCT scans by robustly combining 1) point cloud extraction via 2D segmentation and 2) 3D reconstruction from point clouds via neural fields. Specifically, a 2D segmentation network works as a teacher module to obtain sample points on the airway wall; the neural field then functions as a student module,  receiving and filtering the raw point clouds to obtain an accurate reconstruction of 3D geometry at infinite resolution. 


\subsection{Teacher Module}

Suppose there are $M$ consecutive frames $\{\mathbf{I}_{i=1,...,M}\}$ acquired in a particular OCT scan $\mathcal{I}_k$. $\mathcal{S}_k$ represents the 3D geometry corresponding to $\mathcal{I}_k$ that we wish to extract. A deep neural network  $\Phi(\mathbf{I}_i)$ takes in a rectangular frame, $\mathbf{I}_i$, and then predicts the corresponding segmentation $\mathbf{S}_i$. Suppose $\boldsymbol{p}=(x,y,z)$ is a point on $\mathbf{S}_i$. We then have $\Phi(\boldsymbol{p})=\Phi(\theta, d)$ as the predicted occupancy value for $\boldsymbol{p}$, where 1 indicates being inside and 0 being outside the airway wall. As discussed in Sec.~\ref{sec.prel}, each column in the rectangular frames represents an A-line, which captures the line-of-sight distance from the catheter to the airway wall. We can get the intersecting point coordinates of the A-line and the airway wall by extracting the boundary locations $d_{tiss}$ on the segmentation map. We extract intersecting points in the form of $\mathbf{p}_{tiss}=(\theta, d_{tiss}, t)$. Suppose there are $N$ columns in each frame,  $N$ sample points will be collected for each frame. After processing all frames, we obtain a raw point cloud $\mathcal{P}_k$ (consisting of $M\times N$ points) as an initial representation of $\mathcal{S}_k$ from the OCT scan $\mathcal{I}_k$. 

We then transform $\mathcal{P}_k$ from the cylindrical coordinate system to the Cartesian coordinate system as follows. Suppose the catheter trajectory is a straight line (which is our case) on the z-axis, then we have 
\begin{align}
& \theta=\omega \cdot t \,, r_{tiss} = d_{tiss} \cdot sin({\phi}_{cath}) \,, \\ 
& x = r_{tiss}  \cdot \sin(\theta) \,, y = r_{tiss} \cdot \cos(\theta) \,, z = v_{cath} \cdot t - d_{tiss}(t)cos(\phi_{cath})\,.
\label{eq.polar_to_cast}
\end{align}
We obtain a raw point cloud $\mathcal{P}_k$ represented in the Cartesian coordinate system.

\subsection{Student Module}
Instead of working with raw point clouds, a more direct shape representation is desirable. This could be an explicit mesh, a 3D segmentation represented on a grid, or an implicit level set representation. In \texttt{NeuralOCT} we opt for a neural representation of a level set function as it can easily be fit to a point cloud, provides infinite spatial resolution (as it directly represents the level set \emph{function}), and hence is a convenient representation for downstream shape analysis tasks.


Fig.~\ref{fig.workflow} shows how the student module, $f_k$, captures the airway geometry $\mathcal{S}_k$ from the raw point cloud $\mathcal{P}_k$. $f_k$ is a coordinate-based neural network which takes a point coordinate $\boldsymbol{p}=(x,y,z)$ as its input and outputs a signed distance value $s=f_k(x,y,z)$. The zero-level-set of $f_k$ represents the airway surface $\mathcal{S}_k$. We use the NeuralPull~\cite{neuralpull} loss to obtain a 3D shape reconstruction from the raw point cloud $\mathcal{P}_k$.

In a signed distance function (SDF), we can pull a 3D query location $\boldsymbol{q}_{i}$ to its nearest neighbor $\boldsymbol{t}_{i}$ on the surface using the predicted signed distance $s_{i}$ and the gradient $\boldsymbol{g}_{i}$ at $\boldsymbol{q}_{i}$ within the network, which is the direction of the fastest signed distance increase in 3D space. Therefore, we can use this property to move a query location 
to its nearest point on the surface. Assuming that the SDF is negative on the inside and positive on the outside of the shape 
\begin{equation}
\boldsymbol{t}_{i}^{\prime}=\boldsymbol{q}_{i}-\boldsymbol{f}\left(\boldsymbol{q}_{i}\right) \nabla \boldsymbol{f}\left(\boldsymbol{q}_{i}\right) /\left\|\nabla \boldsymbol{f}\left(\boldsymbol{q}_{i}\right)\right\|_{2}\,,
\label{eq.np_1}
\end{equation}
where $\boldsymbol{t}_{i}^{\prime}$ is the pulled query location $\boldsymbol{q}_{i}$ after pulling, $\nabla \boldsymbol{f}\left(\boldsymbol{q}_{i}\right) /\left\|\nabla \boldsymbol{f}\left(\boldsymbol{q}_{i}\right)\right\|_{2}$ is the direction of gradient $\nabla \boldsymbol{f}\left(\boldsymbol{q}_{i}\right)$. Since $\boldsymbol{f}$ is a continuously differentiable function, we can obtain $\nabla \boldsymbol{f}\left(\boldsymbol{q}_{i}\right)$ in the back-propagation process when training $f_k$. As Fig.~\ref{fig.workflow} illustrates, for query locations inside of the shape $\mathcal{S}$, if the sign of the signed distance value is negative,  the network will move the query location $\boldsymbol{q}_{i}$ along the direction of the gradient to $\boldsymbol{t}_{i}^{\prime}$ on $\mathcal{S}$ using $\boldsymbol{t}_{i}^{\prime}=\boldsymbol{q}_{i}+$ $\left|\boldsymbol{f}\left(\boldsymbol{q}_{i}\right)\right| \nabla \boldsymbol{f}\left( \boldsymbol{q}_{i}\right) /\left\|\nabla \boldsymbol{f}\left(\boldsymbol{q}_{i}\right)\right\|_{2}$. Conversely, the network will move query locations outside of $\mathcal{S}$ against the direction of the gradient due to the positive signed distance value, using $\boldsymbol{t}_{i}^{\prime}=\boldsymbol{q}_{i}-\left|\boldsymbol{f}\left(\boldsymbol{q}_{i}\right)\right| \nabla \boldsymbol{f}\left(\boldsymbol{q}_{i}\right) /\left\|\nabla \boldsymbol{f}\left(\boldsymbol{q}_{i}\right)\right\|_{2}$.

To design a loss to train $\boldsymbol{f}$ so that it represents the point cloud $\mathcal{P}$ the intuition is as follows. For a given query point, $\boldsymbol{q}_i$, the pulled location, $\boldsymbol{t}_i^{\prime}$, results in the closest point on the surface if $\boldsymbol{f}$ is indeed a good signed distance function corresponding to the point cloud $\mathcal{P}$. We then just need to compare $\boldsymbol{t}_i^{\prime}$ to the actual closest point $\boldsymbol{t}_i$ to $\boldsymbol{q}_i$ in the point cloud $\mathcal{P}$. Their difference should be small if $\boldsymbol{f}$ is a good SDF for $\mathcal{P}$. This directly motivates the squared error loss
\begin{equation}
d\left(\left\{\boldsymbol{t}_{i}^{\prime}\right\},\left\{\boldsymbol{t}_{i}\right\}\right)=\frac{1}{I} \sum_{i \in[1, I]}\left\|\boldsymbol{t}_{i}^{\prime}-\boldsymbol{t}_{i}\right\|_{2}^{2} 
\end{equation}
where $I$ is the number of queried points.

\section{Experimental Setting}

\noindent\textbf{Dataset.} Our Airway OCT dataset consists of a total of 35 airway OCT scans. Each OCT scan consists of 100 to 600 frames. 931 frames from the 35 OCT scans have manual airway segmentations. We split the dataset by scans to prevent information leaks from similar frames. The training set consists of 25 OCT scans (819 segmented frames) and the testing set of 10 (112 segmented frames). All of the frames are resized to 1024 $\times$ 1024. The intensities are truncated to $(\mu-\sigma, \mu+\sigma)$ and then are rescaled to $[0,1]$ with min-max normalization.~\footnote{For a frame, $\mu$ is mean intensity and $\sigma$ is the standard deviation.} The OCT scanning parameters are available in the supplementary material.

\vskip1ex
\noindent\textbf{Comparisons.} For 2D OCT segmentation, we compare MedSAM~\cite{ma2024medsam}, nnUNet~\cite{isensee2021nnu}, and NISF~\cite{stolt2023nisf}, which is an implicit segmentation method. To reduce the inference time, we use the MedSAM encoder to produce the latent codes for NISF~\cite{stolt2023nisf} instead of optimizing global latent codes during training and inference. 

\vskip1ex
\noindent\textbf{Evaluations.}
Since there is no 3D ground truth. We are only able to perform quantitative evaluations based on 2D segmentations. We also evaluate qualitatively  by visualizing the extracted raw point cloud and reconstructed meshes. We use segmentation metrics such as the DICE score; geometric metrics: Hausdorff, Chamfer, and earth mover's distances; and the A-line metrics $\mu_{dist}$ and $M_{dist}$, which capture the error in the light-of-sight distance estimations. For each frame, $\mu_{dist}$ is the mean line-of-sight (LOS) distance error:  $\mu_{dist}=\frac{1}{N}\Sigma |d_{tiss}^{gt} - d_{tiss}^{pred}|$;  $M_{dist}$ is the maximum LOS  distance error: $M_{dist}= max_(|d_{tiss}^{gt} - d_{tiss}^{pred}|)$. We also evaluate reconstruction accuracy by measuring the distances from the raw point cloud $\mathcal{P}_k$ to the reconstructed mesh $\mathcal{S}_k$. 

\vskip1ex
\noindent\textbf{Training Details.} We used an NVIDIA 3909 Ti (24GB) GPU for training. The training of the nnUNet took $\sim$ 16h, while the training of the implicit neural model using the NeuralPull loss took $\sim$ 1h for all test scans.

\section{Results}

\begin{table}[]
\vspace{-1cm}
\caption{Quantitative evaluation of 2D OCT segmentation.}
\resizebox{1.0\columnwidth}{!}{%
\begin{tabular}{|l|c|c|c|c|c|c|c|}
\hline
Methods                     & NeuralPull      & CD $\times$ $10^3$ $\downarrow$              & HD $\downarrow$               & EMD $\times$ $10^3$ $\downarrow$                 & DICE $\uparrow$              & $\mu_{dist}$ (mm) $\downarrow$      & $M_{dist}$ (mm) $\downarrow$       \\ \hline
\multirow{2}{*}{MedSAM}     & \XSolidBrush & 2.615 $\pm$ 7.695 & 0.068 $\pm$ 0.100 & 14.581 $\pm$ 28.925 & 0.982 $\pm$ 0.037  & 0.177 $\pm$ 0.350 & 0.974 $\pm$ 1.523 \\
                            & \Checkmark   & 3.175 $\pm$ 9.291 & 0.070 $\pm$ 0.117 & 16.232 $\pm$ 30.259 & 0.981 $\pm$  0.038 & 0.197 $\pm$ 0.366 & 0.957 $\pm$ 1.609 \\ \hline
\multirow{2}{*}{Local-NISF} & \XSolidBrush & 1.353 $\pm$ 4.658 & 0.052 $\pm$ 0.085 & 9.797 $\pm$ 17.494  & 0.989 $\pm$ 0.017  & 0.119 $\pm$ 0.212 & 0.774 $\pm$ 1.339 \\
                            & \Checkmark  & 1.417 $\pm$ 4.845 & 0.050 $\pm$ 0.085 & 10.575 $\pm$ 17.106 & 0.989 $\pm$ 0.016  & 0.128 $\pm$ 0.207 & 0.745 $\pm$ 1.338   \\ \hline
NeuralOCT &
  \XSolidBrush &
  0.306 $\pm$ 1.318 &
  0.037 $\pm$ 0.066 &
  \textbf{4.723} $\pm$ \textbf{6.041} &
  \textbf{0.995} $\pm$ \textbf{0.007} &
  \textbf{0.057} $\pm$ \textbf{0.073} &
  0.527 $\pm$ 0.903 \\
(from nnUNet) &
  \Checkmark &
  \textbf{0.290} $\pm$ \textbf{1.310} &
  \textbf{0.035} $\pm$ \textbf{0.063} &
  5.479 $\pm$ 5.397 &
  0.994 $\pm$ 0.006 &
  0.066 $\pm$ 0.065 &
  \textbf{0.525 $\pm$ 0.883} \\ \hline
\end{tabular}}
\footnotesize{ $^1$ \(\mathrm{CD}=\) Chamfer distance. \(\mathrm{EMD}=\) Earth mover's distance. \(\mathrm{HD}=\) Hausdorff distance. $^2$ A \Checkmark in NeuralPull means the segmentation is derived from reconstructed geometries while a \XSolidBrush means that segmentation from the teacher module is evaluated.}
\vspace{-0.5cm}
\label{tab.2d_seg}
\end{table}

Tab.~\ref{tab.2d_seg} shows that the nnUNet produces the best segmentation results in terms of segmentation accuracy and geometric fidelity, indicating that training from scratch with an nnUNet works better than using a pre-trained foundation model for our OCT image segmentation. We suspect the reason is that although the large-scale training set of MedSAM includes two retina OCT datasets MedSAM does not generalize well to airway OCT images. 

\begin{figure}
\vspace{-0.25in}
    \centering
    \includegraphics[width=1.0\textwidth]{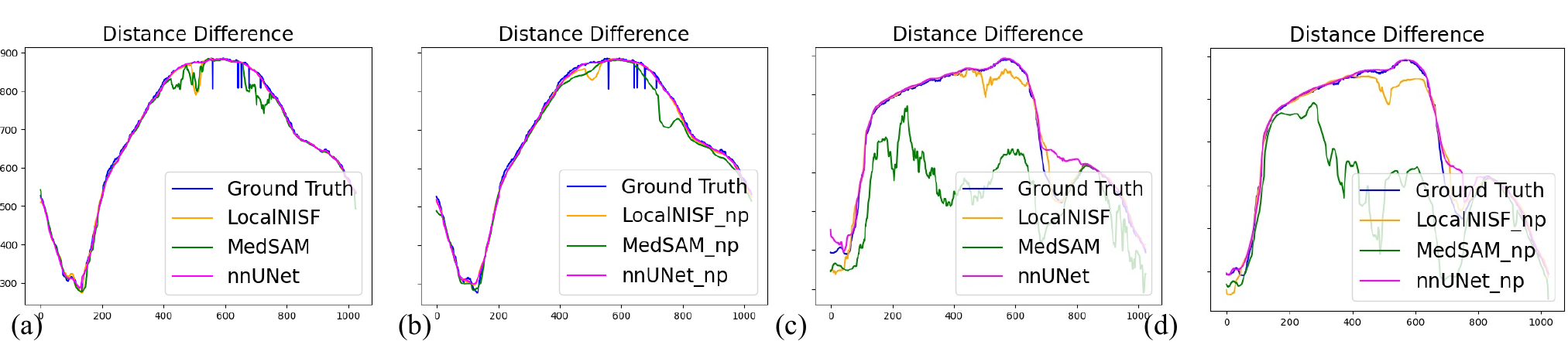}
    \caption{Visualizations of the airway OCT segmentation. (a,c) compares the predicted light-of-sight distances from the teacher module with different methods on scan A and scan B; (b,d) compares the predicted light-of-sight distances from the student module on scan A and scan B. More visualizations are available in the supplementary material.}
    \label{fig.curve}
\end{figure}

\begin{figure}
\vspace{-0.1in}
    \centering
    \includegraphics[width=0.75\textwidth]{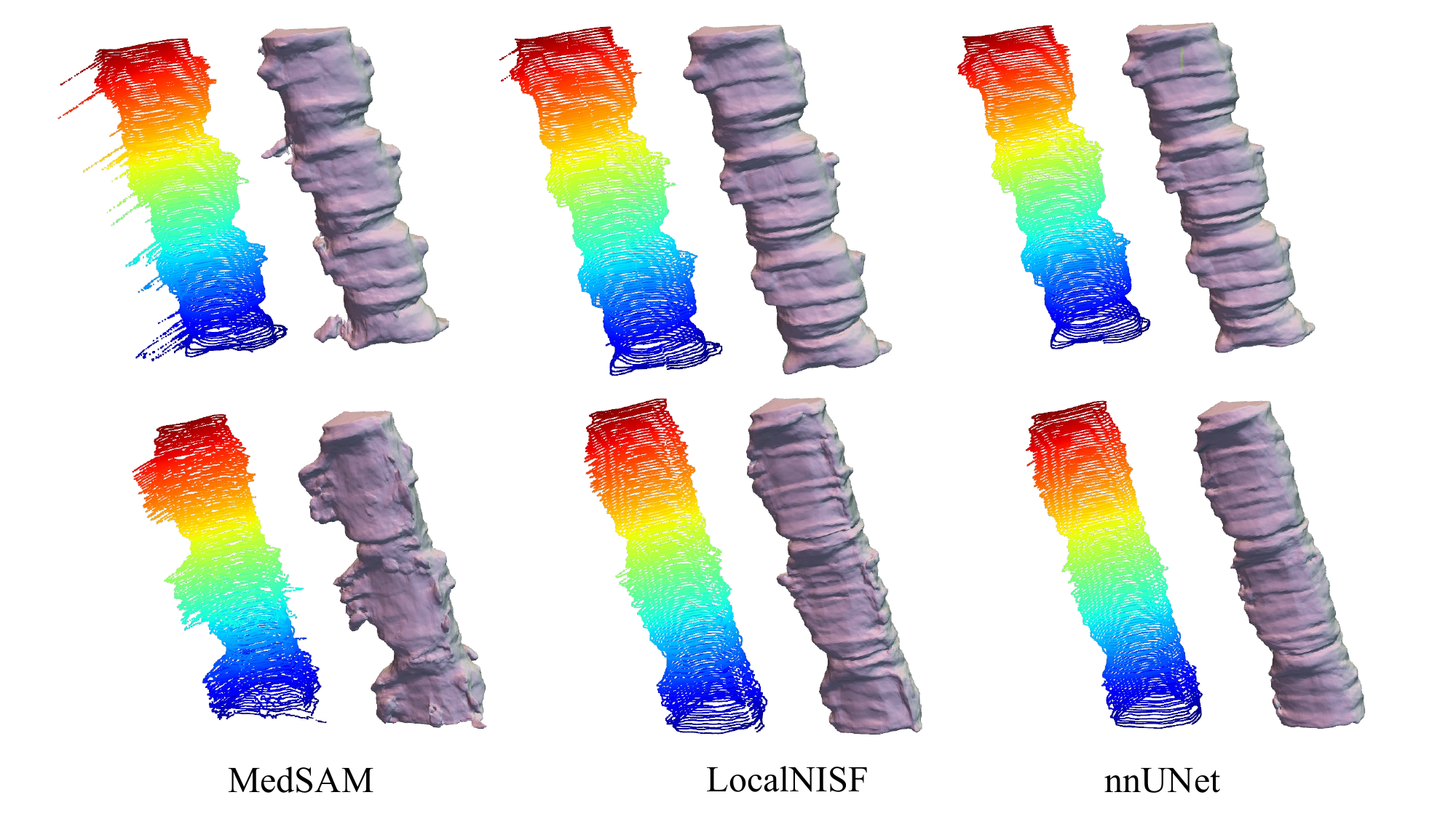}
    \caption{Visualizations of the 3D airway geometry reconstructions. Each pair of shapes represents the raw point clouds from the teacher module and 3D reconstructions from the student module respectively.}
    \vspace{-0.5cm}
    \label{fig.shapes}
\end{figure}

For MedSAM and Local-NISF, the raw segmentations from the teacher module are more accurate than those extracted from neural fields; while for \texttt{NeuralOCT} the 2D segmentations from neural fields are more robust predictions (as captured by a smaller std of metrics) while maintaining similar accuracy. Fig.~\ref{fig.curve} shows that all airway wall reconstructions from neural fields look smoother than those directly obtained from the segmentation network.

\begin{table}[]
\vspace{-0.1in}
\centering
\resizebox{0.7\columnwidth}{!}{%
\begin{tabular}{|l|c|c|c|}
\hline
Methods    & CD    $\times$ $10^3$ $\downarrow$              & HD $\downarrow$         & EMD   $\times$ $10^3$ $\downarrow$             \\ \hline
MedSAM     & 0.049 $\pm$ 0.063 & 0.080 $\pm$ 0.081 & 2.494$\pm$ 1.150 \\
Local NISF & 0.023 $\pm$ 0.015 & 0.046 $\pm$ 0.041 & 1.99 $\pm$0.297  \\
NeuralOCT & \textbf{0.022 $\pm$ 0.013} & \textbf{0.040 $\pm$ 0.029} & \textbf{1.904 $\pm$ 0.255} \\ \hline
\end{tabular}%
}
\caption{Quantitative evaluation of 3D Reconstruction with Neural Fields.}
\vspace{-0.75cm}
\label{tab.3d_rec}
\end{table}

Fig.~\ref{fig.shapes} shows the raw point clouds $\mathcal{P}_k$ from different teacher modules and geometry reconstructions via neural fields. We can observe that the nnUNet produces the visually cleanest raw point clouds, while MedSAM produces the noisiest ones. The 3D reconstructions from neural fields accurately approximate the raw point clouds and tends to eliminate outliers. 

Tab.~\ref{tab.3d_rec} shows the estimation errors between the raw point clouds and the respective 3D reconstructions: \texttt{NeuralOCT} produces the best reconstructions. 



\section{Conclusion}
In this work, we proposed \texttt{NeuralOCT}, a learning-based approach to extract 3D airway geometry from OCT scans. \texttt{NeuralOCT} combines 1) point cloud extraction via 2D segmentation and 2) 3D reconstruction from raw point clouds via neural fields. Our quantitative and qualitative evaluations show that the reconstructed geometries of \texttt{NeuralOCT} are accurate and robust (line-of-distance error < 70 micrometer). Future work may be able to achieve improved reconstructions by jointly optimizing the teacher module and the student module. 

\newpage
\section*{\centering{Acknowledgement}}

The research reported in this publication was supported by NIH grant 1R01HL154429. The content is solely the responsibility of the authors and does not necessarily represent the official views of the NIH.

\clearpage
\bibliography{references}

\begin{thebibliography}{10}
\providecommand{\url}[1]{\texttt{#1}}
\providecommand{\urlprefix}{URL }
\providecommand{\doi}[1]{https://doi.org/#1}

\bibitem{achlioptas2018learning}
Achlioptas, P., Diamanti, O., Mitliagkas, I., Guibas, L.: Learning representations and generative models for {3D} point clouds. In: ICML. pp. 40--49 (2018)

\bibitem{bu2017airway}
Bu, R., Balakrishnan, S., Iftimia, N., Price, H., Zdanski, C., Oldenburg, A.L.: Airway compliance measured by anatomic optical coherence tomography. Biomedical optics express  \textbf{8}(4),  2195--2209 (2017)

\bibitem{drexler2008state}
Drexler, W., Fujimoto, J.G.: State-of-the-art retinal optical coherence tomography. Progress in retinal and eye research  \textbf{27}(1),  45--88 (2008)

\bibitem{gropp2020implicit}
Gropp, A., Yariv, L., Haim, N., Atzmon, M., Lipman, Y.: Implicit geometric regularization for learning shapes. arXiv:2002.10099  (2020)

\bibitem{groueix2018papier}
Groueix, T., Fisher, M., Kim, V.G., Russell, B.C., Aubry, M.: A papier-m{\^a}ch{\'e} approach to learning {3D} surface generation. In: CVPR. pp. 216--224 (2018)

\bibitem{isensee2021nnu}
Isensee, F., Jaeger, P.F., Kohl, S.A., Petersen, J., Maier-Hein, K.H.: nnu-net: a self-configuring method for deep learning-based biomedical image segmentation. Nature methods  \textbf{18}(2),  203--211 (2021)

\bibitem{jiao2023texttt}
Jiao, Y., Zdanski, C.J., Kimbell, J.S., Prince, A., Worden, C.P., Kirse, S., Rutter, C., Shields, B., Dunn, W.A., Mahmud, J., et~al.: {NAISR}: A {3D} neural additive model for interpretable shape representation. In: ICLR (2023)

\bibitem{kirillov2023sam}
Kirillov, A., Mintun, E., Ravi, N., Mao, H., Rolland, C., Gustafson, L., Xiao, T., Whitehead, S., Berg, A.C., Lo, W.Y., et~al.: Segment anything. arXiv:2304.02643  (2023)

\bibitem{neuralpull}
Ma, B., Han, Z., Liu, Y.S., Zwicker, M.: Neural-pull: Learning signed distance functions from point clouds by learning to pull space onto surfaces. arXiv:2011.13495  (2020)

\bibitem{noise2noisemapping}
Ma, B., Liu, Y.S., Han, Z.: Learning signed distance functions from noisy {3D} point clouds via noise to noise mapping  (2023)

\bibitem{ma2024medsam}
Ma, J., He, Y., Li, F., Han, L., You, C., Wang, B.: Segment anything in medical images. Nature Communications  \textbf{15}(1), ~654 (2024)

\bibitem{mescheder2019occupancy}
Mescheder, L., Oechsle, M., Niemeyer, M., Nowozin, S., Geiger, A.: Occupancy networks: Learning {3D} reconstruction in function space. In: CVPR. pp. 4460--4470 (2019)

\bibitem{niemeyer2019occupancy}
Niemeyer, M., Mescheder, L., Oechsle, M., Geiger, A.: Occupancy flow: {4D} reconstruction by learning particle dynamics. In: CVPR. pp. 5379--5389 (2019)

\bibitem{osher2005level}
Osher, S., Fedkiw, R.P.: Level set methods and dynamic implicit surfaces, vol.~1 (2005)

\bibitem{park2019deepsdf}
Park, J.J., Florence, P., Straub, J., Newcombe, R., Lovegrove, S.: Deepsdf: Learning continuous signed distance functions for shape representation. In: CVPR. pp. 165--174 (2019)

\bibitem{price2018geometric}
Price, H.B., Kimbell, J.S., Bu, R., Oldenburg, A.L.: Geometric validation of continuous, finely sampled {3-D} reconstructions from {aOCT} and {CT} in upper airway models. IEEE TMI  \textbf{38}(4),  1005--1015 (2018)

\bibitem{sethian1999level}
Sethian, J.A.: Level set methods and fast marching methods: evolving interfaces in computational geometry, fluid mechanics, computer vision, and materials science, vol.~3 (1999)

\bibitem{sitzmann2020siren}
Sitzmann, V., Martel, J., Bergman, A., Lindell, D., Wetzstein, G.: Implicit neural representations with periodic activation functions. NeurIPS  \textbf{33},  7462--7473 (2020)

\bibitem{stolt2023nisf}
Stolt-Ans{\'o}, N., McGinnis, J., Pan, J., Hammernik, K., Rueckert, D.: Nisf: Neural implicit segmentation functions. In: MICCAI. pp. 734--744 (2023)

\bibitem{tearney2012consensus}
Tearney, G.J., Regar, E., Akasaka, T., Adriaenssens, T., Barlis, P., Bezerra, H.G., Bouma, B., Bruining, N., Cho, J.m., Chowdhary, S., et~al.: Consensus standards for acquisition, measurement, and reporting of intravascular optical coherence tomography studies: a report from the international working group for intravascular optical coherence tomography standardization and validation. Journal of the American College of Cardiology  \textbf{59}(12),  1058--1072 (2012)

\bibitem{wijesundara2014quantitative}
Wijesundara, K., Zdanski, C., Kimbell, J., Price, H., Iftimia, N., Oldenburg, A.L.: Quantitative upper airway endoscopy with swept-source anatomical optical coherence tomography. Biomedical optics express  \textbf{5}(3),  788--799 (2014)

\bibitem{wu2016learning}
Wu, J., Zhang, C., Xue, T., Freeman, B., Tenenbaum, J.: Learning a probabilistic latent space of object shapes via {3D} generative-adversarial modeling. NeurIPS  \textbf{29} (2016)

\bibitem{zhou2023dlforoctseg}
Zhou, Z.Q., Guo, Z.Y., Zhong, C.H., Qiu, H.Q., Chen, Y., Rao, W.Y., Chen, X.B., Wu, H.K., Tang, C.L., Su, Z.Q., et~al.: Deep learning-based segmentation of airway morphology from endobronchial optical coherence tomography. Respiration  \textbf{102}(3),  227--236 (2023)

\bibitem{zhuang2023octpipeline}
Zhuang, Z., Chen, D., Liang, Z., Zhang, S., Liu, Z., Chen, W., Qi, L.: Automatic {3D} reconstruction of an anatomically correct upper airway from endoscopic long range {OCT} images. Biomedical Optics Express  \textbf{14}(9),  4594--4608 (2023)

\end{thebibliography}
\bibliographystyle{splncs04}

\end{document}


\begin{table}
\centering
\caption{Parameters for airway OCT scans.}\label{table.surface_par}
\begin{tabular}{l l l}
\toprule
Parameter & Values & Explanations\\
\midrule
$f_{samp}$& 100kHz& The scanning frequency\\
$v_{cath}$& 0.6 cm/s& The moving velocity of the catheter \\
${\omega}$& 20Hz& The angular velocity of the laser rays \\
${\phi}_{cath}$&75$^\circ$ & The angle between the catheter and the laser rays \\
\bottomrule
\end{tabular}
\end{table}

\begin{figure}
    \centering
    \includegraphics[width=0.8\textwidth]{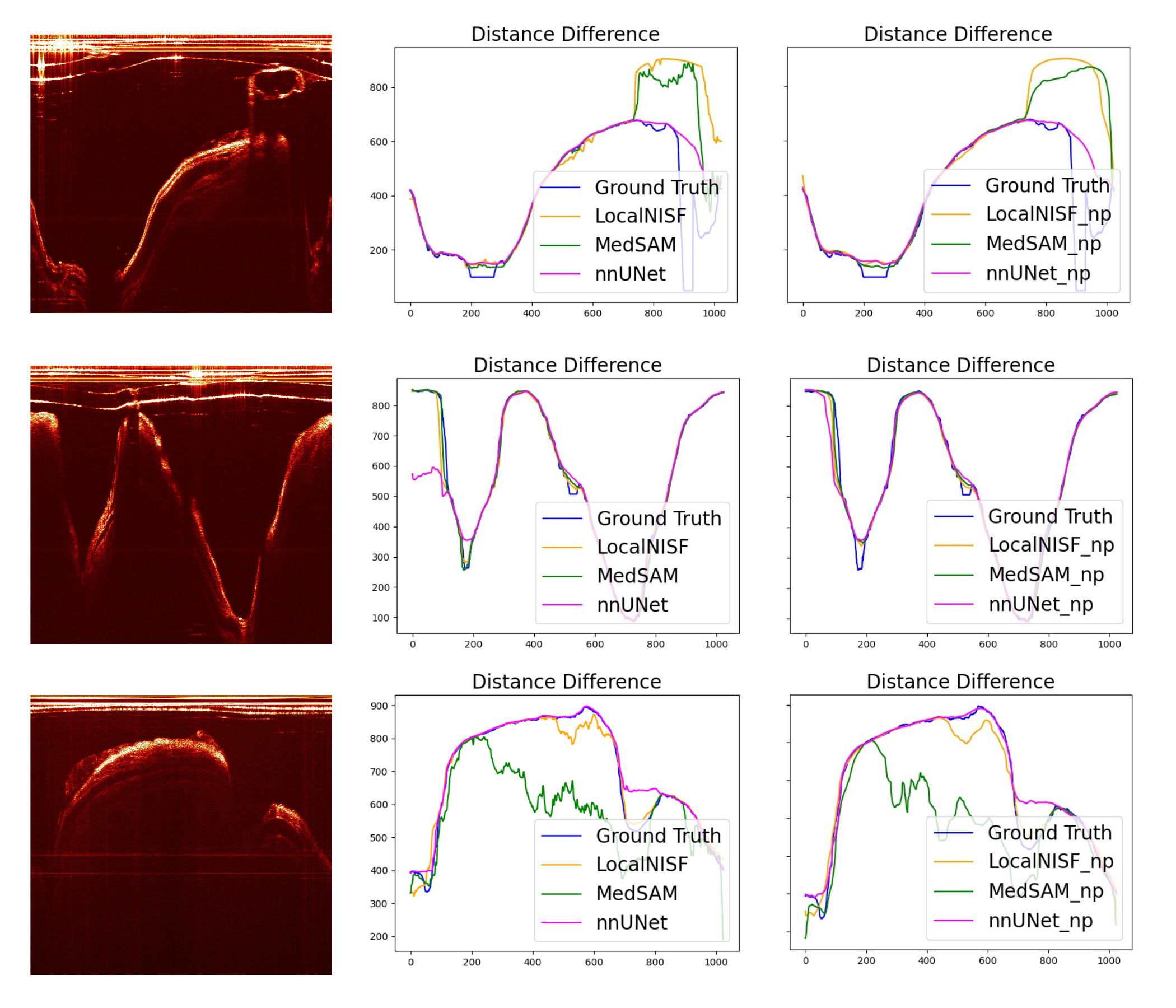}
    \caption{\small Visualizations of the airway OCT segmentation. The first column visualizes OCT frames; the second column compares different teacher modules; the third column compares 2D OCT geometries from the student modules. }
    \label{fig:enter-label}
\end{figure}